\documentclass[referee,sn-nature,oneside]{sn-jnl}

\usepackage{graphicx}%
\usepackage{multirow}%
\usepackage{amsmath,amssymb,amsfonts}%
\usepackage{amsthm}%
\usepackage{mathtools}
\usepackage{mathrsfs}%
\usepackage[title]{appendix}%
\usepackage{xcolor}%
\usepackage{textcomp}%
\usepackage{manyfoot}%
\usepackage{booktabs}%
\usepackage{algorithm}%
\usepackage{algorithmicx}%
\usepackage{algpseudocode}%
\usepackage{listings}%
\usepackage{lineno}
\usepackage{subcaption}
\usepackage{placeins}
\begin{document}

\title[Article Title]{RAFT-UP: Robust Alignment for Spatial Transcriptomics with Explicit Control of Spatial Distortion}

\author[1]{\fnm{Yaqi} \sur{Wu}}

\author[2]{\fnm{Jingfeng} \sur{Wang}}

\author[3]{\fnm{Xin Maizie} \sur{Zhou}}

\author*[1]{\fnm{Yanxiang} \sur{Zhao}}\email{yxzhao@email.gwu.edu}

\author*[2,4]{\fnm{Zixuan} \sur{Cang}\email{zcang@ncsu.edu}}

\affil[1]{\orgdiv{Department of Mathematics}, \orgname{George Washington University}, \orgaddress{\state{Washington D.C}, \country{USA}}}

\affil[2]{\orgdiv{Department of Mathematics}, \orgname{North Carolina State University}, \orgaddress{\city{Raleigh}, \state{NC}, \country{USA}}}

\affil[3]{\orgdiv{Department of Biomedical Engineering}, \orgname{Vanderbilt University}, \orgaddress{\city{Nashville}, \state{TN}, \country{USA}}}

\affil[4]{\orgdiv{Center for Research in Scientific Computation}, \orgname{North Carolina State University}, \orgaddress{\city{Raleigh}, \state{NC}, \country{USA}}}

\abstract{
Spatial transcriptomics (ST) profiles gene expression across a tissue section while preserving the spatial coordinates. Because current ST technologies typically profile two-dimensional tissue slices, integrating and aligning slices from different regions of the same three-dimensional tissue or from samples under different conditions enables analyses that reveal 3D organization and condition-associated spatial patterns. Two major challenges remain. First, interpretable and flexible control over spatial distortion is needed because rigid transformations can be overly restrictive, whereas highly deformable mappings may arbitrarily distort spatial proximity. Second, biologically plausible matching is also needed, especially when the slices overlap partially. Here, we introduce RAFT-UP, a tool for robust ST alignment that provides explicit control over spatial distance preservation through a fused supervised Gromov-Wasserstein (FsGW) optimal transport framework. FsGW combines expression and spatial information, incorporates spot-wise constraints to discourage biologically implausible matches, and enforces a pairwise distance-consistency constraint that prevents mapping two pairs of spots when their spatial distances differ beyond a specified tolerance. We demonstrate that RAFT-UP accurately aligns slices from different regions of the same tissue and slices from different samples. Benchmarking shows that RAFT-UP improves spatial distance preservation while achieving spot label matching accuracy comparable to state-of-the-art methods. Finally, we demonstrate RAFT-UP on two spatially constrained downstream applications, including spatiotemporal mapping of developing mouse midbrain and comparative cross-slice analysis of cell-cell communication. RAFT-UP is available as open-source software.
}

\keywords{Fused Supervised Gromov-Wasserstein, Partial Alignment, Spatial Transcriptomics}

\maketitle

\section*{Introduction}

Spatial transcriptomics (ST) technologies \cite{tian2023expanding,rao2021exploring} measure gene expression across spatially resolved locations within tissue sections. There are several ST technologies, including sequencing-based approaches such as 10x Visium \cite{staahl2016visualization}, Slide-seq \cite{rodriques2019slide}, and Stereo-seq \cite{chen2022spatiotemporal}, which provide transcriptome-wide coverage at different spatial resolutions, and imaging-based approaches such as seqFISH \cite{shah2016situ}, osmFISH \cite{codeluppi2018spatial}, and MERFISH \cite{xia2019spatial}, which achieve subcellular resolution for selected genes. ST data has provided critical insights into cellular organization and tissue architecture in their native spatial context, important information that is lost in single-cell RNA sequencing data (scRNA-seq).
Various computational approaches have been developed specifically for extracting biological knowledge from ST data. These include methods for identifying spatially variable genes \cite{chen2024evaluating}, inferring cell-cell communication processes \cite{armingol2024diversification}, and segmenting spatial tissue domains \cite{yuan2024benchmarking}. These analyses often focus on individual ST slices. 

To gain a comprehensive understanding of a system and to identify key differences across conditions, it is crucial to perform joint analyses of multiple ST slices from different regions of the same three-dimensional tissue or from different samples representing different conditions or stages. Several computational methods have been developed for aligning ST slices. GPSA \cite{jones2023alignment} is a probabilistic method that uses a two-layer Gaussian process where a warping function maps spatial coordinates to a common coordinate system (CCS) and a second function maps CCS to phenotype, that is, gene expression profile. It provides alignment in two modes, mapping both slices to a \textit{de novo} CCS or fixing one slice as the CCS and mapping the other one onto it. PRECAST \cite{liu2023probabilistic} is another probabilistic approach that uses spatial factor analysis to obtain joint latent embeddings of multiple slices. STalign \cite{clifton2023stalign} is a geometric approach based on large deformation diffeomorphic metric mapping (LDDMM), and represents the alignment in the form of a diffeomorphism between the spatial domains of the slices. SPACEL \cite{xu2023spacel} is a comprehensive tool with three modules including Spoint for cell type deconvolution using a multi-layer perception model and a probabilistic model, Splane for domain segmentation across multiple slices using a graph convolution neural network, and subsequently Scube that performs a rigid-body transformation to stack the slices. There are also several deep learning-based approaches including DeepST, SPIRAL, and STAligner, which employ different variants of graph neural network architectures to derive shared latent embeddings of spots. The shared latent embeddings obtained in DeepST \cite{xu2022deepst}, SPIRAL \cite{guo2023spiral}, STAligner \cite{zhou2023integrating}, and PRECAST provide an implicit form of alignment, which can be converted into explicit spot-to-spot correspondences through postprocessing such as k-nearest neighbor matching. Additionally, STAligner and SPIRAL also provide dedicated functionality that extends the shared latent embeddings into three-dimensional reconstructions, where STAligner performs rigid transformation using the iterative closest point algorithm and SPIRAL first performs a fused Gromov-Wasserstein optimal transport to obtain a transport matrix connecting the two slices and maps each spot to the average coordinates of its top corresponding spots in the other slice.

The problem of aligning multiple slices of ST data can be viewed as a special case of integrating and aligning high-dimensional and high-resolution omics datasets. Optimal Transport (OT) \cite{villani2008optimal,peyre2019computational} is naturally suited for such tasks, as it provides a general framework for finding correspondence between complex distributions with the flexibility to account for structural similarity. OT has been widely applied to finding various correspondences among single-cell or spatial omics data. For example, WaddingtonOT \cite{schiebinger2019optimal} and SpaTrack \cite{shen2025inferring} use OT to infer developmental trajectories from time series of scRNA-seq datasets and ST datasets, respectively. SCOT \cite{demetci2022scot} uses Gromov-Wasserstein (GW) \cite{memoli2011gromov} OT to align scRNA-seq data and single-cell ATAC sequencing data, while novoSpaRc \cite{moriel2021novosparc}, SpaOTsc \cite{Cang2020InferringData}, and TACCO \cite{mages2023tacco} use different variants of the fused Gromov-Wasserstein (FGW) OT \cite{vayer2020fused} to integrate scRNA-seq data with spatial gene expression data. Recently, several approaches have utilized OT for aligning ST slices including PASTE \cite{zeira2022alignment} using FGW, PASTE2 \cite{liu2023partial} using partial FGW, and moscot \cite{klein2025mapping} using an FGW algorithm with improved efficiency. The FGW approaches used in these tools utilize gene expression dissimilarity as inter-dataset costs and promote the preservation of intra-dataset spatial distances. In these OT-based approaches, the alignment is represented by the transport plans in the form of probability matrices. PASTE and PASTE2 further generate a consensus slice to which every slice is mapped, by computing the FGW barycenter.

Existing methods either rely on rigid transformations, which lack the flexibility to capture local spatial variation, or impose only soft penalties, which could lead to substantial distortions in spatial distances. Precise control of spatial distance distortion across aligned ST slices is critical for biologically meaningful downstream analyses, particularly in applications with strong spatial constraints such as cell-cell communication analysis \cite{armingol2024diversification}. To address these limitations, we present RAFT-UP, a \underline{R}obust \underline{A}lignment \underline{F}ramework for spatial \underline{T}ranscriptomics with \underline{U}ser-\underline{P}rescribed spatial distortion control. RAFT-UP also avoids biologically implausible mapping and simultaneously infers the partial overlap portion between slices. The core algorithm of RAFT-UP is a new Fused supervised Gromov-Wasserstein (FsGW) optimal transport method. FsGW extends our previous works, supervised OT \cite{cang2022supervised} and supervised GW \cite{cang2025supervised}, to simultaneously incorporate element-wise constraints on the transport matrix to prohibit correspondences between dissimilar spots and higher-order constraints on pairs of elements in the transport matrix to enforce distance preservation within a tolerance. In the FsGW framework, the cross-dataset cost is derived from a joint embedding of the slices, obtained by extending our earlier graph neural network-based spatially aware embedding method \cite{cang2021scan} to train in an alternating scheme across different slices. To handle large datasets, we first downsample the spots to obtain a uniformly spaced set of points. The FsGW is then computed on the downsampled data, after which the full transport matrix is recovered by solving a supervised OT problem.

We show that RAFT-UP accurately aligns ST slices from different regions of the same tissue and from samples across different developmental stages, using real datasets generated by various technologies including 10X Visium, Stereo-seq, and MERFISH. The benchmarking results demonstrate that RAFT-UP effectively manages the trade-off between spatial preservation and accuracy on cell type matching. It outperforms rigid transformation based approaches in accuracy and achieves better spatial proximity preservation than the approaches allowing spatial flexibility while also achieving top accuracy. We further demonstrate the utility of RAFT-UP in comparative analysis of cell-cell communication, a process with strong spatial constraint, across multiple slices.

\section*{Results}
\subsection*{Overview of RAFT-UP}

\begin{figure}[!htbp]
    \centering
    \includegraphics[width=0.97\linewidth]{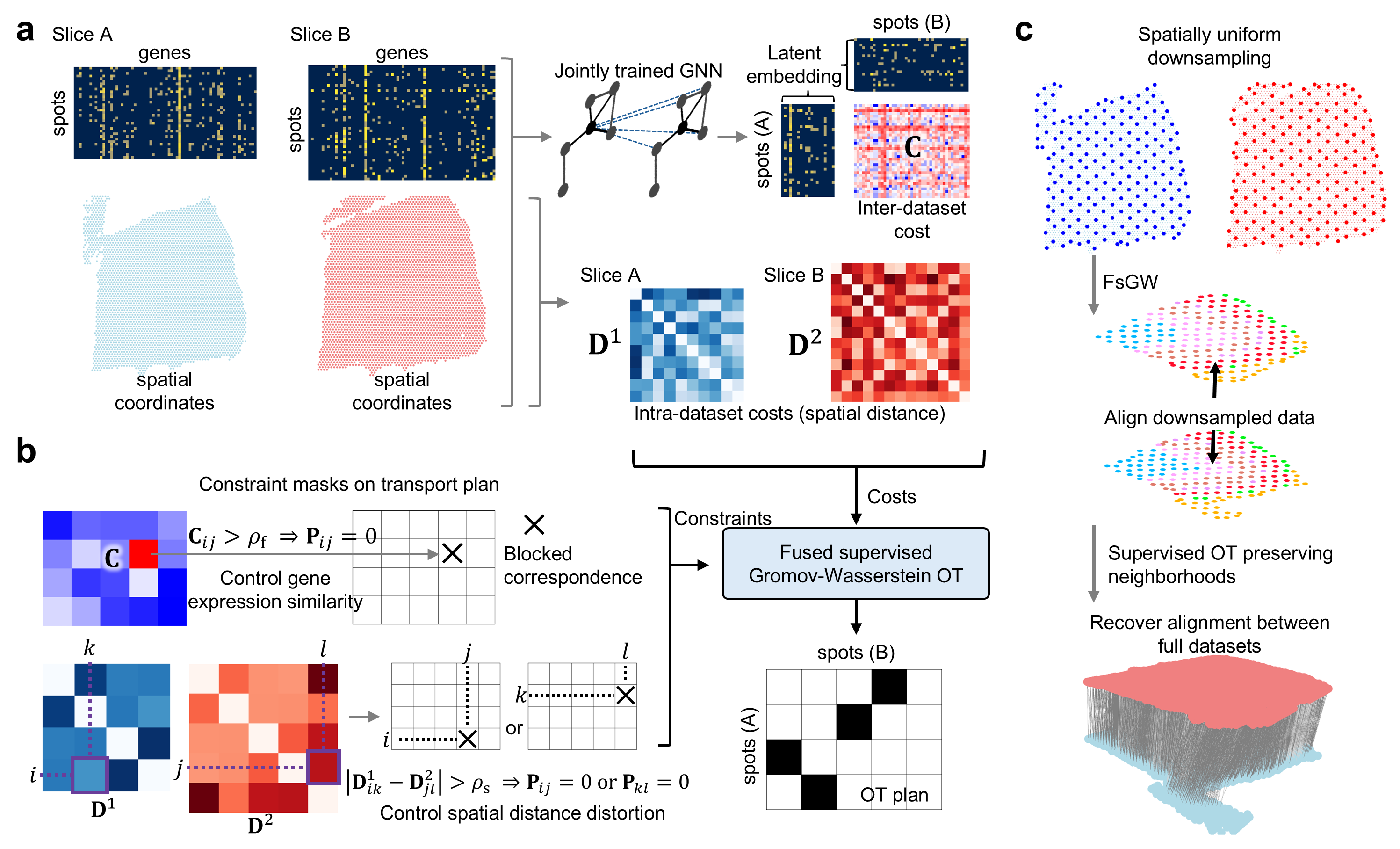}
    \caption{\textbf{Overview of RAFT-UP.} \textbf{a}, An inter-dataset cost matrix that capture gene expression dissimilarity between spots and their spatial neighborhoods is obtained from a Deep Graph Infomax model trained alternately on the two datasets. The two intra-dataset cost matrices are obtained from the spatial distances between the spots within each slice. \textbf{b}, An optimal probabilistic mapping between the two slices is obtained by solving a Fused supervised Gromov-Wasserstein optimal transport, which takes the costs defined in (a) and enforces element-wise and second-order constraints on the transport plan. \textbf{c}, To handle large ST datasets, RAFT-UP uses a downsampling and lifting strategy. An initial mapping is first obtained across geometrically uniform downsampling of the two slices. A full alignment is then recovered by solving a supervised optimal transport problem with a cost based on the distance of each spots to the anchor points in the initial mapping.}
    \label{fig:overview}
\end{figure}

Across spatially adjacent or temporally consecutive ST slices, there is often moderate spatial distortion, variation in gene expression, and partial overlap. We therefore seek an alignment method that both allows these differences in the mapping and regulates their extent. To address these challenges, RAFT-UP uses a Fused supervised Gromov-Wasserstein (FsGW) optimal transport framework. FsGW integrates supervised OT \cite{cang2022supervised} and supervised GW \cite{cang2025supervised}, extending FGW to incorporate element-wise and higher-order constraints on the transport plan, for controlling gene expression similarity and spatial distance preservation, respectively. Unlike the partial OT framework, which requires a predetermined transported mass, FsGW naturally infers the proportion of partial overlap from these constraints. An overview of the core algorithm of RAFT-UP is shown in Fig. \ref{fig:overview}.

Given two ST slices of $n_1$ and $n_2$ spots, an inter-dataset cost $\mathbf{C}\in\mathbb{R}_+^{n_1\times n_2}$ is computed from a joint embedding of the spots in the two slices, which describes the gene expression profile of each spot and its spatial neighborhood. We train a Deep Graph Infomax model \cite{velickovic2018deep,cang2021scan} alternately on the two datasets to obtain the joint embedding, and later compute $\mathbf{C}_{ij}$ as the Euclidean distance of spot $i$ of slice A and spot $j$ of slice B in the embedding space. Two intra-dataset cost matrices $\mathbf{D}^1\in\mathbb{R}_+^{n_1\times n_1}$ and $\mathbf{D}^2\in\mathbb{R}_+^{n_2\times n_2}$ are computed from the spatial distances between the spots within each slice (Fig. \ref{fig:overview}a). RAFT-UP then finds an optimal coupling matrix $\mathbf{P}^*\in\mathbb{R}^{n_1\times n_2}$ between the slices by solving a FsGW problem that has the following properties. (1) $\mathbf{P}^*_{ij}$ has a high weight indicating correspondence between spot $i$ in slice A and spot $j$ in slice B if the two spots' spatial gene expression profiles are similar. (2) RAFT-UP can also explicitly forbid matches whose expression dissimilarity exceeds a user-controlled threshold $\rho_\mathrm{f}$, which helps prevent mismatches such as aligning different cell types. (3) When both $\mathbf{P}^*_{ij}$ and $\mathbf{P}^*_{kl}$ have high values, the intra-dataset distances $\mathbf{D}^1_{ik}$ and $\mathbf{D}^2_{jl}$ are similar. (4) RAFT-UP further explicitly controls the spatial structure preservation by enforcing that any pair of correspondences can only coexist when the distance distortion is within a threshold $\rho_\mathrm{s}$. As a result, RAFT-UP naturally obtains a partial coupling using the interpretable parameters $\rho_\mathrm{f}$ and $\rho_\mathrm{s}$ that control feature consistency and geometry consistency, respectively (Fig. \ref{fig:overview}b). 

In practice, RAFT-UP uses a downsampling, FsGW coupling, and lifting to full data workflow to improve the computational efficiency (Fig. \ref{fig:overview}c). Specifically, a geometrically uniform downsampling is first performed on the two slices. For 10X Visium data, we exploit the near-hexagonal grid of spots. For other ST data, we use a max-min sampling procedure \cite{de2004topological} to promote even spreading of sampled points across the tissue domain. We then compute an FsGW coupling between the two downsampled slices. Finally, we lift this correspondence to the full datasets by constructing a cross-slice spatial cost for all spot pairs, which is derived from their spatial proximity to the sampled points within each slice and the precomputed cross-slice correspondence of the sampled points. Using this cross-slice spatial cost for full data, we solve a supervised optimal transport problem \cite{cang2022supervised} to obtain a coupling over all spots. Details of the RAFT-UP model and algorithm are provided in \textbf{Methods} and \textbf{Supplementary Note}.

\subsection*{Achieving high accuracy while preserving spatial structure}

\begin{figure}[!htbp]
    \centering
    \includegraphics[width=0.88\linewidth]{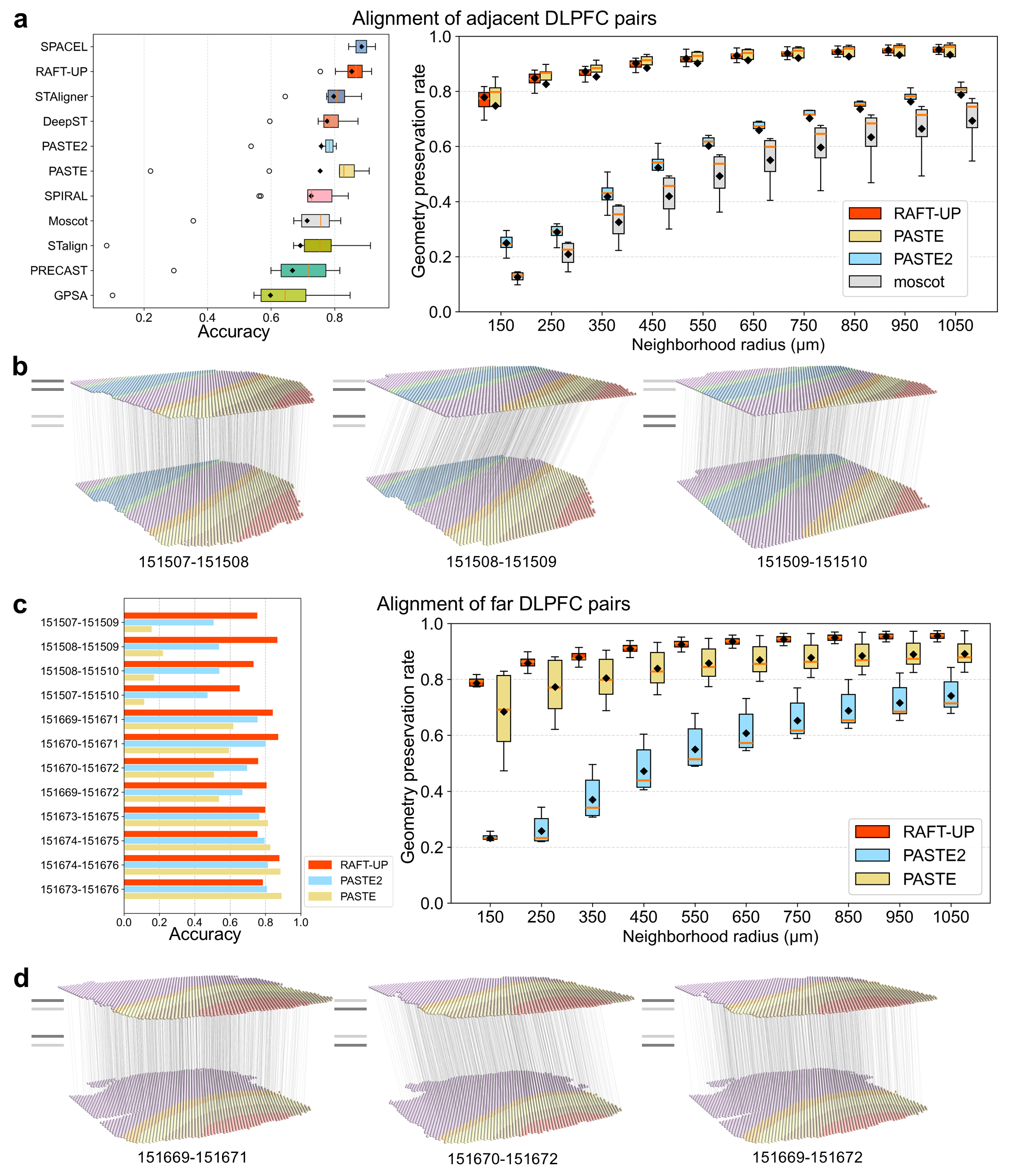}
    \caption{\textbf{Benchmarking alignment on human DLPFC Visium slices.} \textbf{a}, Performance on adjacent DLPFC section pairs. Left: alignment accuracy for RAFT-UP and other methods. Right: geometric preservation rate (GPR) as a function of neighborhood radius, comparing RAFT-UP, PASTE2, PASTE, and moscot. SPACEL is evaluated under its native, label-informed configuration and the remaining methods are evaluated without label information. The accuracy of PRECAST, STalign, DeepST, SPIRAL, and GPSA are taken from a recent benchmark \cite{hu2024benchmarking}. \textbf{b}, Representative RAFT-UP alignments of three adjacent pairs. Spots are colored by annotated cortical layer and line segments indicate the computed spot-level correspondence between slices. \textbf{c}, Performance on non adjacent DLPFC pairs. Left: Accuracy for each pair comparing RAFT-UP, PASTE2, and PASTE. Right GPR versus neighborhood radius for far pairs. \textbf{d}, Representative RAFT-UP alignments for three far pairs. In the boxplots, boxes indicate the interquartile range (25th-75th percentiles) with the median as the center line, whiskers extend to the most extreme values within 1.5×IQR, and black diamonds denote the mean.}
    \label{fig:benchmark_dlpfc}
\end{figure}

\begin{figure}[!htbp]
    \centering
    \includegraphics[width=0.85\linewidth]{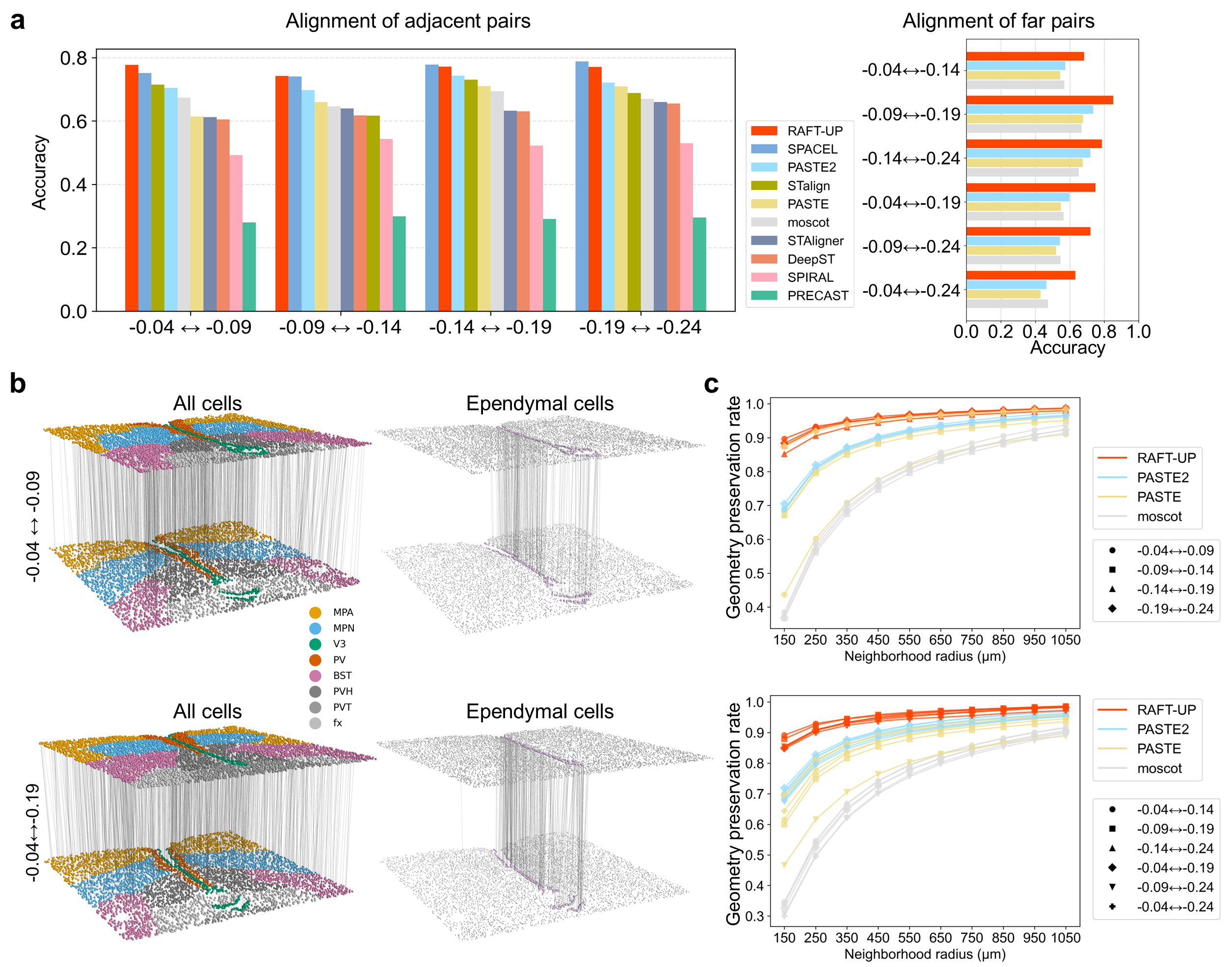}
    \caption{\textbf{Benchmarking alignment on MERFISH sections.} \textbf{a}, Label matching accuracy for MERFISH data for adjacent pairs (left) and far pairs (right). SPACEL is evaluated under its native label-informed configuration and other methods are evaluated without access to label information. \textbf{b}, Representative RAFT-UP alignments for an adjacent pair (top) and a far pair (bottom) showing both alignment of all cells (left) and alignment of ependymal cells (right) highlighting its performance on aligning fine structures. \textbf{c}, Geometric preservation rate (GPR) as a function of neighborhood radius comparing RAFT-UP, PASTE2, PASTE, and moscot on adjacent pairs (top) and far pairs (bottom).}
    \label{fig:benchmark_merfish}
\end{figure}

We first evaluate the performance of RAFT-UP on mapping accuracy and preservation of spatial structure using the human dorsolateral prefrontal cortex (DLPFC) dataset generated with the 10x Visium platform \cite{maynard2021transcriptome}. The DLPFC dataset contains 12 cortical sections with expert annotations of cortical layers 1 to 6 and white matter (WM). These 12 slices are organized into three groups, each containing four spatially consecutive sections (A, B, C, and D) from the same tissue sample. The pairs A-B and C-D are 10 $\mu$m apart and the pair B-C is 300 $\mu$m apart. We use two metrics to evaluate the alignment performance, including (1) alignment accuracy of spatial domains and (2) geometric neighborhood preservation. An alignment between two slices is represented by a mapping matrix $\mathbf{P}\in\mathbb{R}^{n_1\times n_2}$, where $\mathbf{P}_{ij}$ is the matching weight between spot $i$ in slice A and spot $j$ in slice B. First, we report alignment accuracy, defined as the fraction of spots whose highest-weight matching in the other slice has the same layer label. Second, we report the Geometric Preservation Rate (GPR), which quantifies how well local spatial neighborhoods, defined within a fixed radius $r$, are preserved under the alignment $\mathbf{P}$, by measuring how much transported mass from a spot's neighborhood maps into the neighborhoods of its matched spots. Both metrics take values in $[0,1]$, with higher values indicating better alignment. See Methods for detailed definition of these evaluation metrics.

The alignment performance of RAFT-UP on DLPFC sections is summarized in Fig. \ref{fig:benchmark_dlpfc}. Across all adjacent alignments, RAFT-UP achieves consistently high alignment accuracy (Fig. \ref{fig:benchmark_dlpfc}a, b). Among the compared methods, the alignment module of the comprehensive SPACEL package utilizes label information under its native setup, and therefore SPACEL achieves the highest label accuracy. The other methods operate under the same setup without access to label information, and among these methods, RAFT-UP achieves the strongest overall accuracy.  
RAFT-UP also shows strong preservation of local spatial geometry on adjacent pairs when compared to other OT-based approaches (Fig. \ref{fig:benchmark_dlpfc}b, Supplementary Fig. 1), where RAFT-UP performs comparably to PASTE and outperform PASTE2 and moscot across neighborhood radii.
To further evaluate the performance on less similar slices, we consider far slice pairs which include the middle-adjacent pairs that are $300\mu$m apart and the non-adjacent pairs (Fig. \ref{fig:benchmark_dlpfc}c, d). RAFT-UP maintains high label accuracy as slice similarity decreases, and achieves the highest average accuracy among the three OT-based methods (RAFT-UP: 0.792, PASTE2: 0.680, PASTE: 0.528, Fig. \ref{fig:benchmark_dlpfc}c), demonstrating robust performance beyond aligning highly similar pairs. RAFT-UP also provides the best geometry preservation on these far pairs (Fig. \ref{fig:benchmark_dlpfc}d, Supplementary Fig. 2), highlighting the benefit of the supervised Gromov-Wasserstein component in maintaining spatial structure when aligning less similar slices.

We next evaluate RAFT-UP on a MERFISH dataset. On adjacent section pairs, RAFT-UP achieves consistently high label matching accuracy across all pairs outperforming all other methods that do not use label information (Fig. \ref{fig:benchmark_merfish}a, Supplementary Fig. 3). On more distant non-adjacent pairs, RAFT-UP maintains high accuracy and substantially improves over unconstrained full mapping and partial OT methods (Fig. \ref{fig:benchmark_merfish}a). Qualitative correspondences further support these quantitative evaluations. RAFT-UP produces coherent cross-section mappings at the whole-tissue level and preserves the spatial localization of low-abundant populations such as ependymal cells across adjacent and farther-separated pairs (Fig. \ref{fig:benchmark_merfish}b, Supplementary Fig. 3). Finally, RAFT-UP consistently achieves better geometric preservation than traditional OT-based approach over a wide range of neighborhood radii for both adjacent and far pairs (Fig. \ref{fig:benchmark_merfish}c), with the largest gains at small radii which is most sensitive to local distortions. Together, these results demonstrate that RAFT-UP preserves spatial structures across various scales while maintaining high label alignment accuracy.

\FloatBarrier

\subsection*{Interpretable alignment of partially overlapping slices}

\begin{figure}[!htbp]
    \centering
    \includegraphics[width=0.6\linewidth]{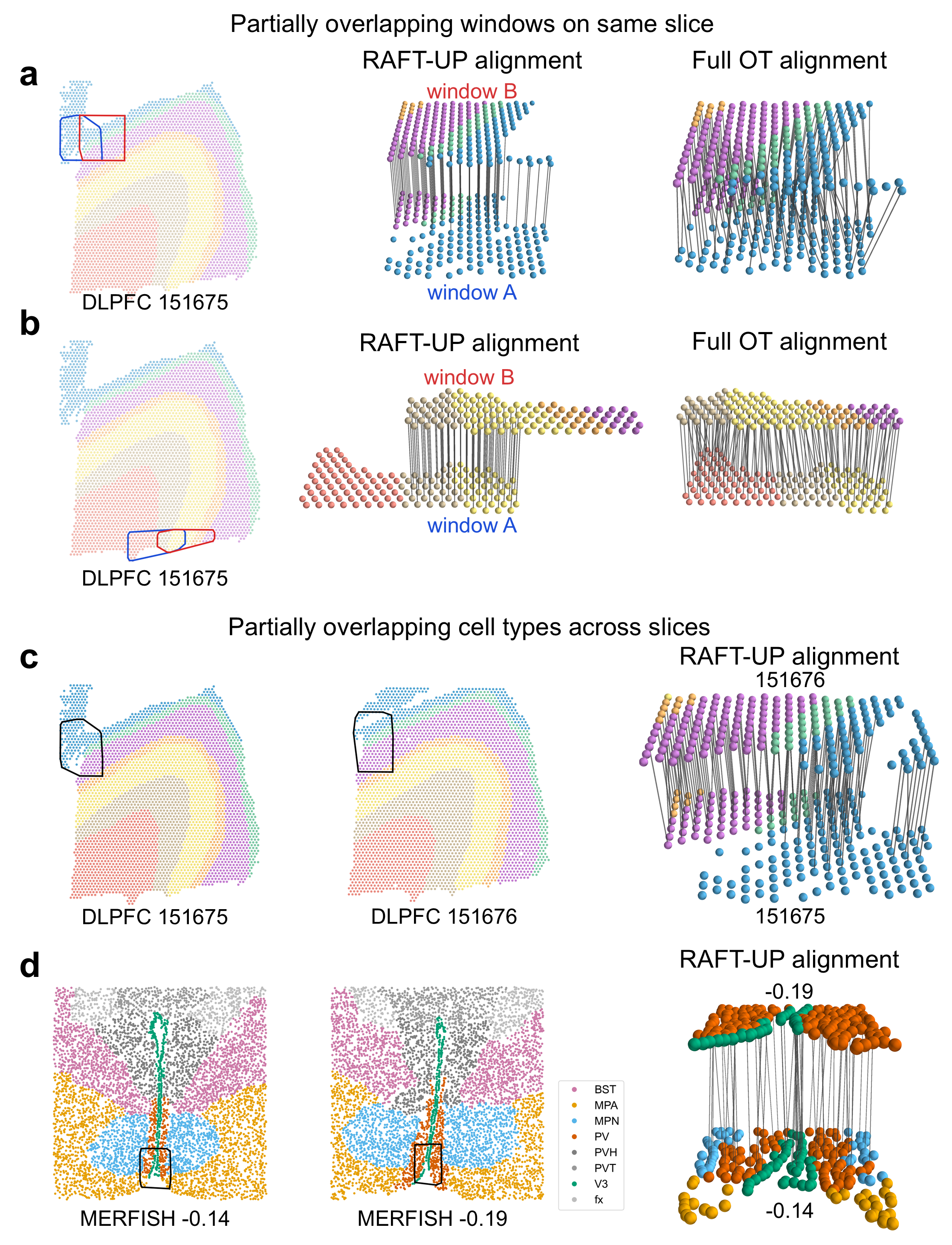}
    \caption{\textbf{Evaluation of RAFT-UP under partial overlap.} \textbf{a, b}, Partially overlapping windows extracted from the same DLPFC slice. RAFT-UP alignment only contain alignment in the truly overlapping region while traditional full OT alignment forces full matching of the two partially overlapping windows. \textbf{c}, Partially overlapping regions with different geometry and cell-type composition across an adjacent DLPFC pair. \textbf{d}, A MERFISH example with a similar geometry but different anatomical region compositions. RAFT-UP aligns cells from anatomical regions present in both windows, while leaving cells from regions absent in one slice largely unmapped.}
    \label{fig:small_windows}
\end{figure}

In practice, particularly when aligning ST slices from different biological replicates, the slices may only partially overlap due to variation in tissue coverage, tearing, or cropping, and their cell-type composition may also vary. Many alignment formulations assume near-complete overlap or require the degree of overlap to be specified, which can lead to spurious correspondences when substantial non-overlapping regions are present. RAFT-UP mitigates this limitation by introducing two interpretable user-controlled parameters, a minimum gene expression similarity threshold $\rho_{\mathrm{f}}$ and a tolerance $\rho_{\mathrm{s}}$ for spatial distance distortion. These together enable robust alignment in partial-overlap scenarios. To demonstrate this utility, we construct several controlled partial-overlap examples with known correspondences at spot level or at cell type level.

First, we extract pairs of partially overlapping windows from the same DLPFC slice and treated them as two inputs for alignment (Fig. \ref{fig:small_windows}a, b). In these within-slice experiments, the main challenge is the partial overlap of spatial regions. RAFT-UP produces correspondences concentrated in the shared region and does not force extensive matches for the non-overlapping portions. In contrast, traditional OT with full matching enforces global mass matching and therefore matches non-overlapping portions, which could also cause large spatial distortion if the geometries are different. Notably, RAFT-UP achieves the correct partial overlapping without requiring a pre-specified overlap proportion.

We further illustrate partial overlap across different slices with different geometry or composition. For a pair of DLPFC slices, RAFT-UP obtains coherent mappings in the region where laminar structure is shared while avoiding forced correspondence in regions lacking clear anatomical counterparts in the other slice (Fig. \ref{fig:small_windows}c). A similar behavior is observed in the MERFISH example, where RAFT-UP aligns cells belonging to anatomical regions present in both windows and leaves the MPA and MPN cells in slice -0.14 largely unmapped as these regions are absent in slice -0.19 (Fig. \ref{fig:small_windows}d). These results suggest that RAFT-UP provides interpretable alignments in the settings with unknown and potentially small overlap.

\FloatBarrier

\subsection*{Modeling spatiotemporal trajectory from ST slices across time points}

\begin{figure}[!htbp]
    \centering
    \includegraphics[width=0.8\linewidth]{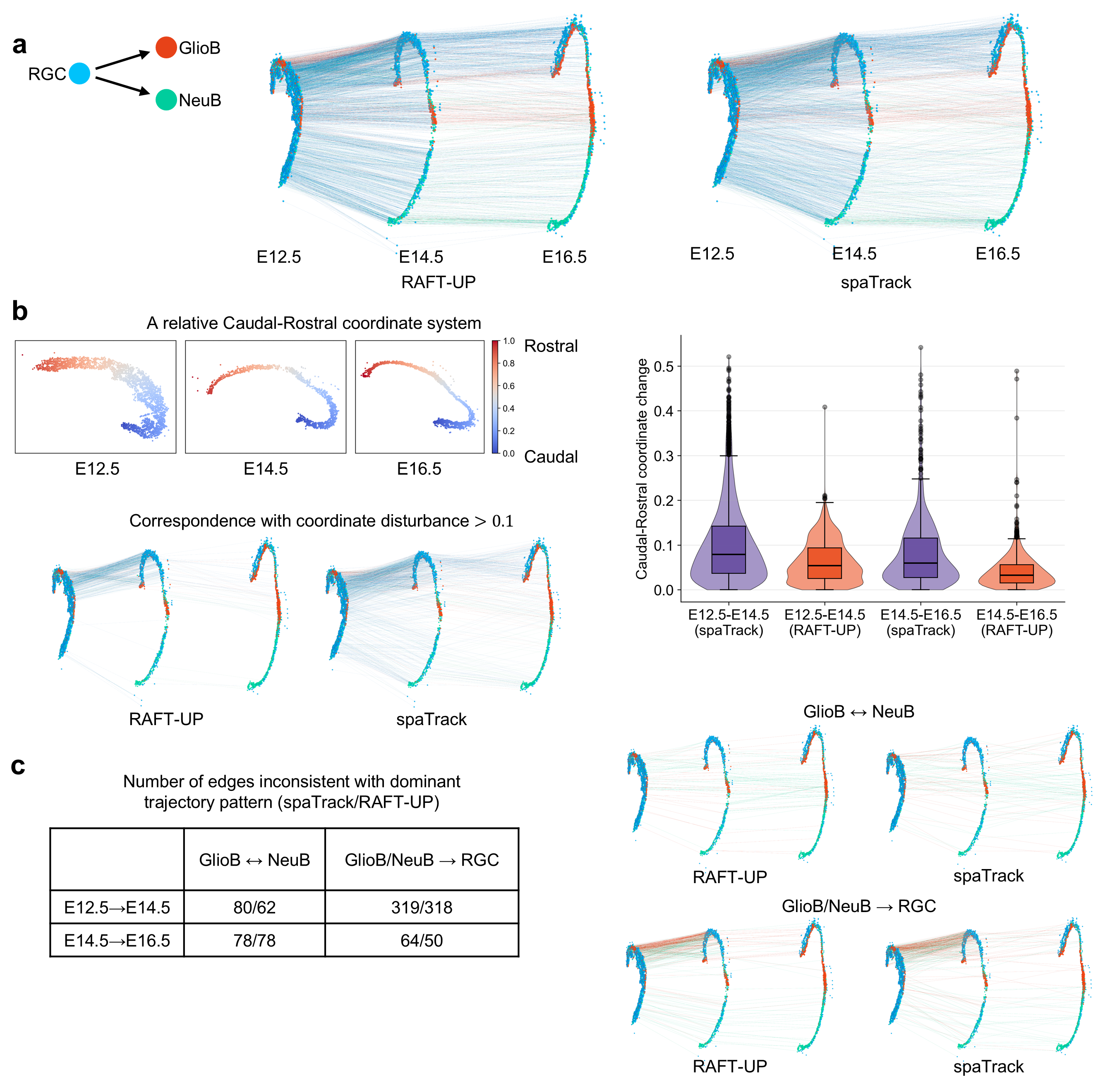}
    \caption{\textbf{Spatiotemporal analysis of mouse midbrain development.} \textbf{a}, Alignment of the Stereo-seq data of mouse midbrain from E12.5 to E14.5 and from E14.5 to E16.5. Cells are colored by the expert annotations from original study: RGC, radial glia cell; GlioB, glioblast; NeuB, neuroblast. \textbf{b}, The disturbance of the relative location of cells on the Caudal to Rostral axis through alignment. The median values of the box plots are 0.0793, 0.0542, 0.0596, 0.0325, from left to right respectively. Cell correspondences with a change of the relative coordinate greater than 0.1 are shown. \textbf{c}, The counts and visualization of correspondences in the alignments that are inconsistent with the dominant trajectory patterns which are RGC to GlioB, RGC to NeuB, or unchanging.}
    \label{fig:spatiotemporal}
\end{figure}

Next, we apply RAFT-UP to Stereo-seq data of mouse midbrain profiled at E12.5, E14.5, and E16.5 to model spatiotemporal trajectories across developmental stages. The dataset includes expert annotations of three major cell states including radial glial cells (RGC), glioblasts (GlioB), and neuroblasts (NeuB), with expected developmental trajectory from RGC to GlioB and NeuB. We use RAFT-UP to perform pairwise alignments between consecutive time points and visualize the inferred cross-time correspondence among the cells (Fig. \ref{fig:spatiotemporal}a). The resulting mapping links cells across time points while preserving their spatial organization within each slice. We quantitatively examine how the inferred correspondences preserve the large-scale spatial ordering along the caudal-rostral axis. For each slice, we define a normalized caudal-rostral coordinate and compute the change in this coordinate for each mapped cell after aligning to the next time point. Across both time pairs, RAFT-UP produces small coordinate shifts (Fig. \ref{fig:spatiotemporal}b). Additionally, correspondences with large coordinate changes ($>0.1$) are relatively rare and are mainly concentrated in regions with substantial shape evolution. 
We also examine consistency between inferred correspondences and the major developmental patterns suggested by the original study. Correspondences of the form RGC to GlioB, RGC to NeuB and same-type matches are treated as consistent, and we count edges that do not fall into these categories (Fig. \ref{fig:spatiotemporal}c).
The RAFT-UP mappings show relatively sparse inconsistent cell correspondences. 

We then compare these results with those obtained using SpaTrack \cite{shen2025inferring}. RAFT-UP exhibits smaller disturbances along the caudal-rostral axis for both time pairs, particularly with much fewer outliers, indicating improved preservation of within-slice spatial arrangement (Fig. \ref{fig:spatiotemporal}b). At the cell state level, RAFT-UP also produces fewer correspondences that are inconsistent with the major developmental transitions (Fig. \ref{fig:spatiotemporal}c). Together, these results suggest that RAFT-UP produces cross-time correspondences that are more spatially preserved and more consistent with expected developmental structure.

\FloatBarrier

\subsection*{Spatially preserving analysis of cell-cell communication across slices}

\begin{figure}[!htbp]
    \centering
    \includegraphics[width=0.9\linewidth]{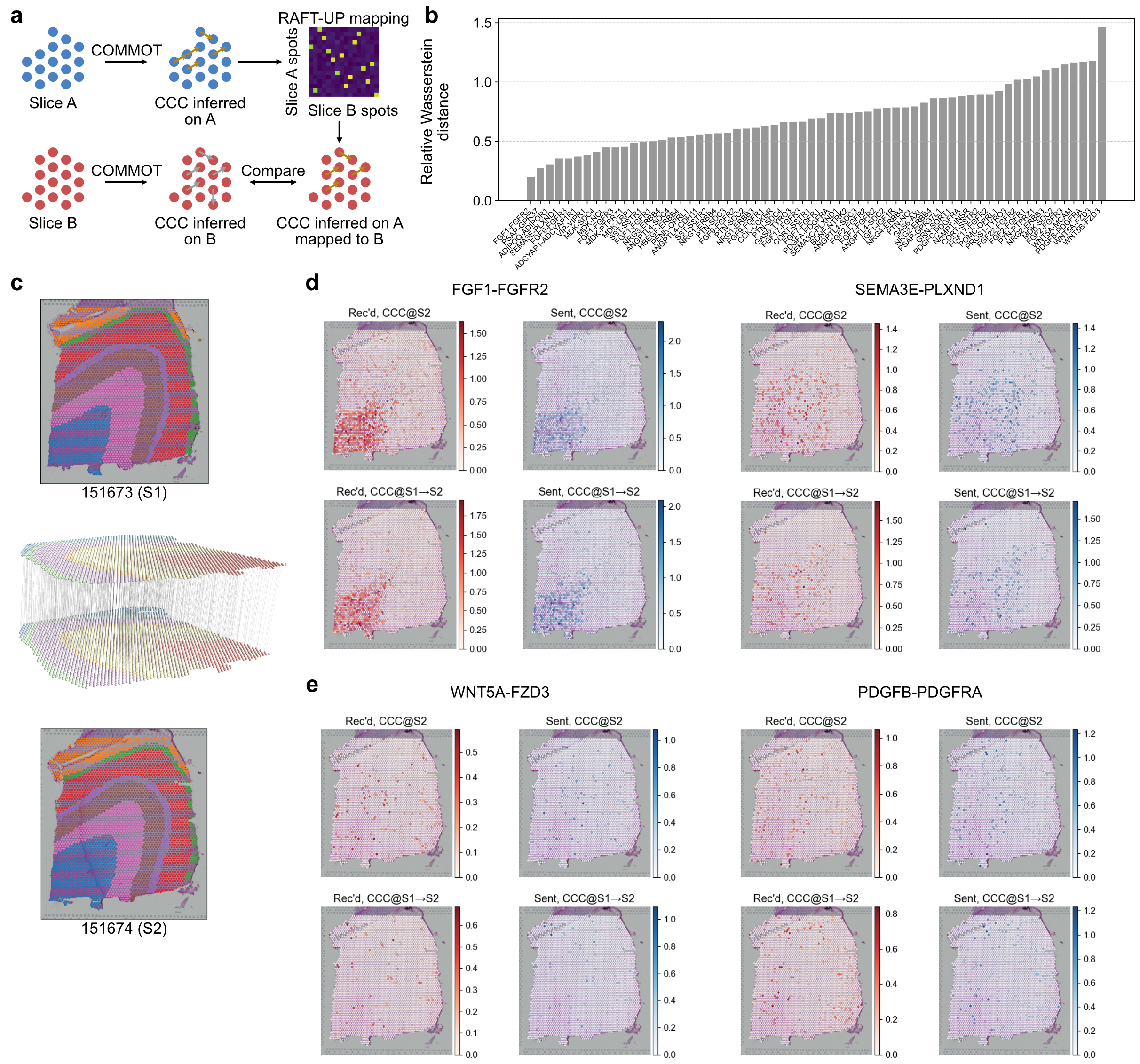}
    \caption{\textbf{Comparative analysis of cell-cell communication.} \textbf{a}, COMMOT is applied independently to each slice to infer cell-cell communication (CCC). CCC inferred on slice A is transferred to slice B using the RAFT-UP spot-to-spot mapping, enabling direct comparison to CCC inferred on slice B, and vice versa. \textbf{b}, Ligand--receptor pairs are ranked by their relative Wasserstein distance between CCC inferred directly on S2 and CCC inferred on S1 mapped onto S2 (CCC@S1$\rightarrow$S2), as well as between CCC inferred directly on S1 and CCC inferred on S2 mapped onto S1 (CCC@S2$\rightarrow$S1) \textbf{c}, RAFT-UP alignment between the two slices where lines indicate matched spots. \textbf{d}, Examples of ligand-receptor pairs with concordant spatial patterns across slices (FGF1-FGFR2 and SEMA3E-PLXND1), shown for received (Rec'd) and sent (Sent) CCC inferred on each slice and mapped in both directions. \textbf{e}, Examples of ligand-receptor pairs with different patterns across slices (WNT5A-FZD3 and PDGFB-PDGFRA).}
    \label{fig:ccc}
\end{figure}

Cell-cell communication (CCC) is important for proper functions of tissues and organs. Because CCC is inherently constrained by spatial proximity, analyses that compare communication patterns across tissue slices are particularly sensitive to geometric distortions introduced during alignment. RAFT-UP is therefore well suited for this task, as it enables explicit control on the preservation of spatial proximity among cells through the alignment (Supplementary Fig. 4). Here, we use RAFT-UP to comparatively analyze CCC across tissue slices. 

We first independently infer CCC on each DLPFC slice using COMMOT \cite{cang2023screening}, a spatially aware CCC inference tool for ST data. For each ligand-receptor (LR) pair, COMMOT outputs a spot-wise directed network in which the weight on edge $(i,j)$ represents the inferred strength of signaling from spot $i$ to spot $j$. Using the transport plan obtained by RAFT-UP from slice A to slice B, we map these communication fields across slices in both directions. Specifically, receiver and sender fields inferred on slice A are transported onto slice B through the transpose of the row-normalized transport plan, while fields inferred slice B are transported to slice A through the column-normalized transport plan. (Fig. \ref{fig:ccc}a).
For each LR pair, we quantify agreement between a mapped field and the corresponding field inferred on the target slice using a relative Wasserstein dissimilarity. For example, to compare the receiver field mapped from slice A to Slice B,  $\mathbf{R}_{\mathrm{A}\rightarrow\mathrm{B}}$, with the one inferred directly on slice B, $\mathbf{R}_\mathrm{B}$, we compute 
\begin{equation}
    d_{\mathrm{relW}}(\mathbf{R}_{\mathrm{A}\rightarrow\mathrm{B}},\mathbf{R}_{\mathrm{B}})=\frac{W_1(\mathbf{R}_{\mathrm{A}\rightarrow\mathrm{B}},\mathbf{R}_\mathrm{B})}{\operatorname{median}_{k}\, W_1(\widetilde{\mathbf{R}}^k_{\mathrm{A}\rightarrow\mathrm{B}},\mathbf{R}_\mathrm{B})},
\end{equation}
where $\widetilde{\mathbf{R}}^k_{\mathrm{A}\rightarrow\mathrm{B}}$ are the mapped fields using a randomly permuted RAFT-UP map, and $W_1$ is the $1$-Wasserstein distance. This normalization addresses the tendency of raw Wasserstein distances to increase for sparse fields and better reveals structural agreements between fields. The LR pairs are then ranked using this dissimilarity score (Fig. \ref{fig:ccc}b,c).

Among the most conserved LR pairs, FGF1-FGFR2 shows a clear spatially structured pattern that is highly preserved after mapping, with both sender and receiver fields enriched in the white matter regions (Fig. \ref{fig:ccc}d, Supplementary Fig. 5). This matches prior knowledge that FGF receptor signaling in oligodendrocytes regulates myelin growth and myelin sheath thickness, where FGFR2 was implicated as a key receptor in this process \cite{furusho2012fibroblast}. 
The LR pair PSAP-GPR37 is also consistent across the slices in the same white matter region (Supplementary Fig. 6), which is consistent with evidence that prosaposin/prosaptide can signal through GPR37-family receptors and that GPR37 influences oligodendrocyte differentiation and central nervous system myelination \cite{meyer2013gpr37}. 
Another highly consistent LR pair is SEMA3E-PLXND1 (Fig. \ref{fig:ccc}d, Supplementary Fig. 7). This signaling pathway has a known function of spatial organization with a documented role in forebrain axon tract development and circuit wiring specificity \cite{chauvet2007gating}. The LR pair WNT5A-FZD3 has also been implicated in spatial organization during cortical wiring \cite{li2009wnt5a}. In contrast to the highly consistent SEMA3E-PLXND1 pattern, WNT5A-FZD3 is substantially less consistent across the two slices (Fig. \ref{fig:ccc}e, Supplementary Fig. 8). This difference may indicate that these guidance related signaling pathways can work in different spatial scales with SEMA3E-PLXND1 forming a more coherent pattern while WNT5A-FZD3 appears more spatially localized and thus less consistent across adjacent slices.
PDGFB-PDGFRA also exhibits a high dissimilarity across the two slices (Fig. \ref{fig:ccc}e, Supplementary Fig. 9). The CCC fields are broadly distributed, but their strongest regions occur in different locations. PDGFRA is a canonical marker of oligodendrocyte precursor cells that persist in the adult brain and continue to generate oligodendrocytes, and PDGF signaling regulates oligodendrocyte development\cite{rivers2008pdgfra}. This suggests that PDGF signaling remains broadly active in adult DLPFC tissue, but its spatial communication field is less organized across adjacent slices.

\FloatBarrier

\section*{Discussion}

We introduced RAFT-UP, an optimal-transport framework for aligning spatial transcriptomics (ST) slices when explicit correspondences are unavailable and the extent of overlap is unknown. In contrast to approaches that rely on rigid transformations or produce only latent-space agreement followed by post hoc matching, RAFT-UP directly outputs an explicit probabilistic coupling between spots. Methodologically, RAFT-UP balances transcriptomic similarity with spatial structure through a fused supervised Gromov-Wasserstein (FsGW) formulation that (1) blocks implausible spot pairings using a feature cutoff ($\rho_\mathrm{f}$) and (2) constrains spatial distance distortion through an interpretable tolerance ($\rho_\mathrm{s}$). These constraints naturally yield a semi-balanced solution that can leave unmatched mass when regions do not overlap, avoiding the need to pre-specify an overlap fraction. To scale to large slices, RAFT-UP employs a downsample-align-upsample workflow: an FsGW coupling is computed on geometrically uniform landmarks and then lifted to full resolution via a supervised OT step.

Across datasets spanning different technologies and biological systems, RAFT-UP consistently achieves a favorable trade-off between matching accuracy and geometric preservation. On Visium DLPFC sections, RAFT-UP attains high layer-wise alignment accuracy while maintaining strong neighborhood preservation, and the benefits are most pronounced for non-adjacent slice pairs where unconstrained transport plans are more likely to introduce long-range crossings or many-to-one assignments. On MERFISH sections, RAFT-UP similarly achieves strong label concordance while preserving fine structures, including low-abundance populations. Controlled partial-overlap experiments on windowed subregions show that RAFT-UP concentrates mass in the truly shared area and avoids forced matches in non-overlapping tissue, unlike full-mass alignment approaches.

An important advantage of RAFT-UP is its explicit and interpretable control over spatial distance preservation during alignment. This is particularly valuable for analyses in which spatial proximity is mechanistically constrained, such as spatio-temporal development of cells and cell-cell communication. We illustrated this advantage in two studies. In the Stereo-seq mouse midbrain developmental series, RAFT-UP produced cross-time correspondences while maintaining global spatial ordering of cells. In the DLPFC cell-cell communication analysis, the geometry-preserving alignment enabled reliable cross-slice comparison of the inferred signaling fields by limiting artificial changes in spatial range and neighborhood structure.  

RAFT-UP does require several user-specified parameters, but the main controls are directly
interpretable and tend to behave consistently across datasets once they are calibrated to platform resolution and the expected biological differences between slices. The feature cutoff $\rho_\mathrm{f}$ determines how strictly potential correspondences must agree in gene expression, whereas the spatial tolerance $\rho_\mathrm{s}$ controls the allowable distortion of within-slice distances. The mixing weight $\alpha$ provides a continuous way to balance transcriptomic similarity against geometric consistency. As in other constrained alignment settings, excessively stringent thresholds may exclude true correspondences and lead to under-alignment, while overly permissive choices can admit spurious links, especially when the slices differ by anisometric deformations or when the learned feature representation is affected by noise.

Several directions may further extend the scope of RAFT-UP. On the algorithmic side, faster solvers and multiscale implementations could improve scalability to very large, densely sampled slices, and joint alignment of multiple slices could enable direct three-dimensional reconstruction without relying on combining pairwise results. From a statistical perspective, incorporating uncertainty quantification for the transport plan would strengthen downstream comparisons and support more reliable hypothesis testing. Finally, integrating RAFT-UP with dynamical optimal transport models offers a natural path toward continuous spatiotemporal inference while explicit controlling geometric distortion, which is crucial in analysis with strong spatial constraints, including communication, migration, and niche dynamics.

\section*{Methods}

Full derivations, optimization algorithms, and implementation details for the optimal transport solvers including fused supervised Gromov-Wasserstein and supervised OT, are provided in the Supplementary Information.

\subsection*{RAFT-UP model}

We consider two spatial transcriptomics (ST) slices, A and B, with $n_1$ and $n_2$ spots, respectively. RAFT-UP represents an alignment between slices in the form of a nonnegative coupling matrix $\mathbf{P}^*\in\mathbb{R}^{n_1\times n_2}$, where $\mathbf{P}^*_{ij}$ represents the matching weight between spot $i$ in slice A and spot $j$ in slice B. Given (1) a cross-slice cost matrix $\mathbf{C}\in\mathbb{R}^{n_1\times n_2}$ describing spatial gene expression dissimilarity between spots across the slices and (2) intra-slice distance matrices $\mathbf{D}^1\in\mathbb{R}^{n_1\times n_1}$ and $\mathbf{D}^2\in\mathbb{R}^{n_2\times n_2}$ quantifying spatial distances between spots within the same slice, RAFT-UP obtains the mapping $\mathbf{P}^*$ by solving the following fused supervised Gromov-Wasserstein (FsGW) optimal transport problem.
\begin{equation}\label{eq:fsgw}
\mathbf{P^*} = \operatorname*{arg\,min}_{\mathbf{P} \in \mathbf{U}(\leq \mathbf{a}, \leq \mathbf{b};\mathbf{C},\mathcal{M})} \;
 \alpha \langle \mathbf{C}, \mathbf{P} \rangle_F
 + (1-\alpha) \langle \mathcal{M}, \mathbf{P} \otimes \mathbf{P} \rangle_F
 + \gamma
 \left(
 \| \mathbf{a}-\mathbf{P}\mathbf{1}_{n_2} \|_{1}
 + \| \mathbf{b}-\mathbf{P}^{\mathsf{T}}\mathbf{1}_{n_1} \|_{1}
 \right),
 \end{equation}
\begin{equation}\label{eq:fsgw_feasible}
\begin{aligned}
 \mathbf{U}(\leq \mathbf{a}, \leq \mathbf{b};\mathbf{C},\mathcal{M}) = \{\mathbf{P}\in\mathbb{R}_+^{n_1\times n_2} \ \big| \ &\mathbf{P}\mathbf{1}_{n_2}\leq \mathbf{a}, \mathbf{P}^\mathsf{T}\mathbf{1}_{n_1}\leq\mathbf{b}, \\
 &\mathbf{P}_{rt}=0\,\,\text{if}\,\,\mathbf{C}_{rt}>\rho_\mathrm{f}, \mathbf{P}_{ij}\mathbf{P}_{kl}=0\,\,\text{if}\,\,|\mathbf{D}^1_{ik}-\mathbf{D}^2_{jl}|>\rho_\mathrm{s}\}. 
\end{aligned}
\end{equation}
Here, $\mathcal{M}$ represents the disturbance of spatial distances through the alignment with $\mathcal{M}_{ijkl}=|\mathbf{D}^1_{ik}-\mathbf{D}^2_{jl}|^2$. The weight parameter $\alpha$ controls the relative contributions of the cross-slice cost ($\langle\mathbf{C},\mathbf{P}\rangle_F$) and disagreement between intra-slice structure ($\langle\mathcal{M},\mathbf{P}\otimes\mathbf{P}\rangle$). For the feasible set (Eq.~\ref{eq:fsgw_feasible}), the first two conditions $\mathbf{P}\mathbf{1}_{n_2}\leq\mathbf{a},
\mathbf{P}^{\mathsf{T}}\mathbf{1}_{n_1}\leq\mathbf{b}$ prevent the transport plan from creating mass, while allowing unmatched mass so the alignment can accommodate only partially overlapping slices. The third constraint $\mathbf{P}_{rt}=0$ if $\mathbf{C}_{rt}>\rho_\mathrm{f}$ prohibits transport between spot pairs whose feature dissimilarity exceeds the cutoff $\rho_\mathrm{f}$. In practice, when $\mathbf{C}$ captures biological differences well, choosing $\rho_\mathrm{f}$ according to cell type separation can largely prevent cell type mismatches. The last constraint controls the extent to which the geometry is preserved. Specifically, it prohibits simultaneous assigning positive masses to $\mathbf{P}_{ij}$ and $\mathbf{P}_{kl}$ if the difference between the within-space distances $\mathbf{D}^1_{ik}$ and $\mathbf{D}^2_{jl}$ exceeds the threshold $\rho_\mathrm{s}$. The third term in Eq.~\ref{eq:fsgw}, $\| \mathbf{a}-\mathbf{P}\mathbf{1}_{n_2} \|_{1}
 + \| \mathbf{b}-\mathbf{P}^{\mathsf{T}}\mathbf{1}_{n_1} \|_{1}$, encourages a more extensive coupling by penalizing untransported mass. This is useful because, under the combined constraints, a fully mass-preserving plan may not exist, and the total transported mass is therefore determined implicitly by the optimization. When dealing with alignment of spatiotemporal data, to account for the spatial growth, a relative spatial preservation constraint is used instead: $\mathbf{P}_{ij}\mathbf{P}_{kl}=0$ if $|\mathbf{D}^1_{ik}-\mathbf{D}^2_{jl}|>\rho_\mathrm{s}\min\{\mathbf{D}^1_{ik},\mathbf{D}^2_{jl}\}$.
 Together, RAFT-UP infers an alignment between slices that balances gene-expression agreement with preservation of intra-slice spatial structure, while automatically determining the coupled mass. The extent of feature and geometric consistency is controlled by the interpretable thresholds $\rho_\mathrm{f}$ and $\rho_\mathrm{s}$. Details of algorithms for solving the FsGW problem and their derivation are in Supplementary Note.

\subsection*{Data preprocessing and construction of cost matrices}

We evaluate our method on the DLPFC Visium dataset (12 slices in total), the MERFISH dataset (5 slices), and the mouse embryo Stereo-seq dataset (3 slices), as well as synthetic windowed slices derived from these datasets. The gene cost matrix for synthetic windowed slices from the same slice is computed from PCA embeddings with 50 PCs using Euclidean distance. For mouse embryo Stereo-seq data, we adopt the gene cost matrix from SpaTrack \cite{shen2025inferring}, which was constructed using the squared Euclidean distance of PCA embeddings with 50 PCs. For DLPFC Visium data, we first filter out genes with total counts fewer than 50 and then identify spatially variable genes by running SOMDE (with 5 nodes per dimension) \cite{hao2021somde} independently on each slice and retaining the top 3000 ranked genes. For MERFISH, we used all measured genes. After gene selection, each slice was normalized by total counts with a target of 10000 counts, log1p-transformed, and scaled (zero mean and unit variance). When aligning a pair of slices, we further filter the genes by keeping genes that are present in both preprocessed slices. 

A joint graph neural network is then used to generate comparable low-dimensional features for the two slices to be aligned for DLPFC Visium datasets and MERFISH dataset. For each slice, a spatial neighborhood graph is initiated as an alpha complex $1$-skeleton \cite{edelsbrunner1994three}, with the radius set to the average distance to the spots' $k$ nearest neighbors ($k$=15). The graph is then densened to increase local connectivity by connecting nodes within three hops in the initial graph yielding the final spatial graph $G_{\alpha}$. Full definition of alpha complex and graph construction details are in Supplementary Note.

Next, we use a graph convolutional encoder $f_\theta$ (shared across the two slices) to map spot features to $h$-dimensional embeddings. 
$f_\theta$ consists of two GCN \cite{kipf2016semi} layers $f_\theta^{(1)},\,f_\theta^{(2)}$ with PReLU activations
\begin{equation}
    H^{(i)}=f_\theta^{(i)}(H^{(i-1)},G)=\mathrm{PReLU}(\tilde{D}^{-\frac12}\tilde{A}\tilde{D}^{-\frac12}H^{(i-1)}W^{(i-1)}), \,\,\, i=1,2,
\end{equation}
where $H^{(0)}$ is the input features, $\tilde{A}$ is the adjacency matrix of the spatial graph $G$ including self connections, and $\tilde{D}$ is the diagonal degree matrix with $\tilde{D}_{ii}=\sum_j\tilde{A}_{ij}$. Both hidden layers are set to 100 dimensions and we take $Z\coloneq H^{(2)}$ as the embedding on which we compute the cross-slice costs. The GCN is trained using the Deep Graph Infomax (DGI) framework \cite{velickovic2018deep}, jointly across the two slices. Let $X\in\mathbb{R}^{n\times d}$ be the $d$-dimensional inputs of the $n$ nodes, DGI constructs a corrupted view of the original graph by permuting node features $\tilde{X}=\pi(X)$ where $\pi$ is a random permutation on rows of a matrix. Let $\{\mathbf{h}_i\}_{i=1}^n=f_\theta(X, G)$ and $\{\tilde{\mathbf{h}}_i\}_{i=1}^n=f_\theta(\tilde{X}, G)$ be the node embeddings from with the original inputs and permutated inputs, respectively. DGI then constructs a global summary of the original graph $\mathbf{s}=\sigma(\frac{1}{n}\sum_i\mathbf{h}_i)$ where $\sigma$ is sigmoid function, and minimizes/maximizes the concordance between node embeddings of perturbed/unperturbed graph with the summary $\mathbf{s}$ with the following loss function
\begin{equation}
    \mathcal{L}_{\mathrm{DGI}}(X,G;\theta,W)=-\frac{1}{n}\sum_i\left[\log D_W(\mathbf{h}_i,\mathbf{s})+\log(1-D_W(\tilde{\mathbf{h}}_i,\mathbf{s}))\right],
\end{equation}
where $D_W(\mathbf{h},\mathbf{s})=\sigma(\mathbf{h}^\mathsf{T}W\mathbf{s})$ is a discriminator with trainable $W$. The GCN is then trained by minimizing the total losses of slice A and slice B 
\begin{equation}
    \min_{\theta,W} \mathcal{L}_{\mathrm{DGI}}(X_\mathrm{A},G_\mathrm{A};\theta,W)+\mathcal{L}_{\mathrm{DGI}}(X_\mathrm{B},G_\mathrm{B};\theta,W).
\end{equation}
We use Adam optimizer with 3500 epochs and learning rate of $0.0002$. After training, we obtain the spot embeddings $Z_\mathrm{A}=f_\theta(X_\mathrm{A},G_\mathrm{A}), Z_\mathrm{B}=f_\theta(X_\mathrm{B},G_\mathrm{B})$, and we use the Euclidean distance between the embeddings as the cross-slice gene-expression cost $\mathbf{C}$.

\subsection*{Downsampling}
We use two downsampling strategies, one for Visium data to exploit the near-hexagonal grid and one for general ST data, with both ensuring even coverage of the slice. For Visium data with spot locations $\mathbf{X}$, a coarse hexagonal grid mask with points $\mathbf{X}^\mathrm{c}$ is first constructed according to a target grid size, typically chosen as an integer multiple of the grid spacing in original data. The coarse grid points are then reduced to a subset $\mathbf{X}_*^{\mathrm{c}}\subset\mathbf{X}^\mathrm{c}$ by keeping those that are the nearest neighbor of at least one point in $\mathbf{X}$. Finally, we choose downsampled spots $\mathbf{X}^*\subset\mathbf{X}$ by including the points in $\mathbf{X}$ that are nearest neighbors of points in $\mathbf{X}_*^\mathrm{c}$. Here, the nearest neighbors are only consider across the two point sets, $\mathbf{X}$ and $\mathbf{X}^\mathrm{c}$.

For other ST data that resembles irregular grids, we use a max-min downsampling approach following \cite{de2004topological}, which is designed to promote even spacing among the downsampled points while retaining extremal points to better respect the geometry. Specifically, the downsampled set is initiated by randomly selecting a point $x_0\in\mathbf{X}$, forming $\mathbf{X}_*^{(0)}=\{x_0\}$. Then points are added to the collection iteratively by selecting 
\begin{equation}
x_i=\operatorname*{arg\,max}_{x_j\in\mathbf{X}\setminus\mathbf{X}_*^{(i-1)}}\left(\min_{x_k\in\mathbf{X}_*^{(i-1)}}D(x_j,x_k)\right),
\end{equation}
where $D$ is the Euclidean distance of the spatial locations of spots. The downsampled set is then updated to $\mathbf{X}_*^{(i)}=\mathbf{X}_*^{(i-1)}\cup\{x_i\}$. This process is repeated until a predefined number of downsampled points is reached.

\subsection*{Recovery of full mapping}
Once, we obtain the alignment matrix between the downsampled slices, we reconstruct the alignment between the full slices by utilizing the relative location of all spots to downsampled spots (anchors) and the correspondence between the anchors. Let $\mathbf{P}\in\mathbb{R}_+^{N_1\times N_2}$ be the optimal transport between the subsampled slices each with $N_1$ and $N_2$ subsampled spots. To avoid fuzzy alignment, we first binarize $\mathbf{P}$ by filtering out all zero rows and columns, replacing the largest entry of each row by $1$ and changing all other entries to $0$, resulting in a $\hat{N}_1\times\hat{N}_2$ binary matrix $\hat{\mathbf{P}}$. We then compute two distance matrices $\hat{\mathbf{D}}^1\in\mathbb{R}_+^{n_1\times \hat{N}_1}$, $\hat{\mathbf{D}}^2\in\mathbb{R}_+^{n_2\times \hat{N}_2}$, describing the spatial distances between the $n_1$ and $n_2$ spots in full slices to the $\hat{N}_1$ and $\hat{N}_2$ anchor spots in slice A and slice B, respectively. These two distance matrices represents the location of each spot relative to the anchor spots within each slice. We further use $\hat{\mathbf{P}}$ to find such relative location across slices. This is achieved by computing
\begin{equation}
    \hat{\mathbf{D}}^{1\rightarrow 2}=\hat{\mathbf{D}}^1\hat{\mathbf{P}}_\mathrm{row}\in\mathbb{R}^{n_1\times \hat{N}_2}_+,\,\,\, \hat{\mathbf{D}}^{2\rightarrow 1}=\hat{\mathbf{D}}^2\hat{\mathbf{P}}_\mathrm{col}^\mathsf{T}\in\mathbb{R}^{n_2\times \hat{N}_1}_+,
\end{equation}
where $\hat{\mathbf{P}}_\mathrm{row}$ and $\hat{\mathbf{P}}_\mathrm{col}$ are row and column normalized $\hat{\mathbf{P}}$ with each row or column summing to $1$. When the $i$th row of $\hat{\mathbf{D}}^{1\rightarrow 2}$ and the $j$th row of $\hat{\mathbf{D}}^2$ are similar, spot $i$ in slice A and spot $j$ in slice B have similar locations under the coordinate system represented by the anchor spots in slice B and therefore should have a small alignment cost. We therefore construct the full cross-slice matrices using
\begin{equation}
    (\mathbf{C}_1)_{ij} = 
\left(
\sum_{k \in \mathcal{K}^1_i}
\left|
\hat{\mathbf{D}}^1_{ik}
-
\hat{\mathbf{D}}^{2\rightarrow 1}_{jk}
\right|^{2}
\right)^{\frac{1}{2}},\,\,
(\mathbf{C}_2)_{ij}
=
\left(
\sum_{l \in \mathcal{K}^2_j}
\left|
\hat{\mathbf{D}}^{1\rightarrow 2}_{il}
-
\hat{\mathbf{D}}^2_{jl}
\right|^{2}
\right)^{\frac{1}{2}},
\end{equation}
Here $\mathcal{K}^1_i = \mathrm{TopK}\!\left(\hat{\mathbf{D}}^1(i,:), k_1\right),
\mathcal{K}^2_j = \mathrm{TopK}\!\left(\hat{\mathbf{D}}^2(j,:), k_2\right)$ focuses on only local anchors to avoid the disruption by long-range spatial distances,
where $\mathrm{TopK}(\cdot,k)$ returns the indices of the $k$ smallest values in a vector corresponding the $k$-nearest-neighbour of a given spot. Finally, we use $\mathbf{C}_\mathrm{t}=(\mathbf{C}_1+\mathbf{C}_2)/2$ as a cross-slice spatial cost relying on the optimal alignment between the subsampled slices. Combined with the cross-slice feature cost $\mathbf{C}$ (as described in Data preprocessing and construction of cost matrices), we solve the following supervised optimal transport to obtain the alignment between full slices
\begin{equation}
    \mathbf{P}^*=\operatorname*{arg\,min}_{\mathbf{P}\in \mathbf{U}(\leq\mathbf{a},\leq\mathbf{b};\mathbf{C},\mathbf{C}_\mathrm{t})}\langle\mathbf{P},\alpha\mathbf{C}+(1-\alpha)\mathbf{C}_\mathrm{t}\rangle_F-\epsilon H(\mathbf{P})+\gamma(\|\mathbf{a}-\mathbf{P}\mathbf{1}_{n_2}\|_1+\|\mathbf{b}-\mathbf{P}^\mathsf{T}\mathbf{1}_{n_1}\|_1),
\end{equation}
\begin{equation}
\begin{aligned}
     \mathbf{U}(\leq \mathbf{a}, \leq \mathbf{b};\mathbf{C},\mathbf{C}_\mathrm{t}) = \Big\{\mathbf{P}\in\mathbb{R}_+^{n_1\times n_2} \ \Big| \  
    &\mathbf{P}\mathbf{1}_{n_2}\leq \mathbf{a}, \mathbf{P}^\mathsf{T}\mathbf{1}_{n_1}\leq\mathbf{b},\\
    &\mathbf{P}_{rt}=0\,\,\text{if }\,\mathbf{C}_{rt}>\rho_\mathrm{f},\\
    &\mathbf{P}_{ij}=0\ \text{if}\ (i,j) \in \mathcal{I}_{\rho_\mathrm{t}}
    \Big\},
\end{aligned}    
\end{equation}
where 
%%%%%
\[
\mathcal{I}_{\rho_\mathrm{t}}
\coloneqq \Big\{(i,j) | \exists k\in\mathcal{K}_i^1,\ 
|\hat{\mathbf{D}}^1_{ik}-\hat{\mathbf{D}}^{2\to 1}_{jk}|>\rho_\mathrm{t}
\ \text{or}\ 
\exists l\in\mathcal{K}_j^2,\ 
|\hat{\mathbf{D}}^{1\to 2}_{il}-\hat{\mathbf{D}}^2_{jl}|>\rho_\mathrm{t}
\Big\},
\]
$\alpha$ is default to $0.9$, $\rho_\mathrm{f}$ is the cross-slice feature cutoff over $\mathbf{C}$ and $\rho_\mathrm{t}$ is the cross-slice spatial cutoff over $\mathbf{C}_\mathrm{t}$ in this recovery step. See Supplementary Note for details of supervised optimal transport algorithms.

\subsection*{Evaluation metrics}

To evaluate alignment quality, we use an alignment accuracy, which quantifies how well aligned
spots preserve cortical layer identity across slices. Let $\mathbf{P} \in \mathbb{R}^{n_1 \times n_2}$ denote the alignment matrix
between slice A and slice B, where $\mathbf{P}_{ij}$ represents the matching
weight between spot $i$ in slice A and spot $j$ in slice B. For each spot $i$ in slice A, we define a hard assignment
\begin{equation*}
j^*(i) = \arg\max_{j} \mathbf{P}_{ij},
\end{equation*}
the alignment accuracy is then defined as
\begin{equation}
\frac{1}{n_1}
\sum_{i}
\mathbf{1}_{
\ell_\mathrm{A}(i) = \ell_\mathrm{B}(j^*(i))},
\end{equation}
where $\mathbf{1}$ denotes the indicator function and $\ell$ is the layer label of the spot.

In addition to alignment accuracy, we introduce a neighborhood preservation metric, called global Geometric Preservation Rate (GPR), to assess the spatial performance of the alignment matrix $\mathbf{P}$. The intuition is that if spot $i$ in slice A is aligned with spot $j$ in slice B, then the $k$-nearest neighbors of $i$ should map consistently to the $k$-nearest neighbors of $j$. This metric provides a way to evaluate whether local geometric structures are preserved under the alignment. Formally, we first define the set of valid indices in slice A
\[
V = \left\{ i \;\middle|\; \sum_{j=1}^{n_2} \mathbf{P}_{ij} > \epsilon_\mathrm{mass} \right\},
\]
which ensures that only spots with sufficient alignment mass are considered. 
For each valid spot $i \in V$, we identify its matched set of spots in slice B,
\[
V_i = \left\{ j \;\middle|\; \mathbf{P}_{ij} > \epsilon_\mathrm{entry} \right\}.
\]

\noindent The preservation rate of spot $i$ is then defined as a weighted average over its 
matches $j \in V_i$:  
\[
\text{preservation-rate}[i] = 
\frac{\sum_{j \in V_i} \mathbf{P}_{ij} \cdot \text{pair-preservation-rate}[ij]}
     {\sum_{j \in V_i} \mathbf{P}_{ij}} ,
\]
where the pairwise preservation rate compares the neighborhood of $i$ with that of $j$:  
\[
\text{pair-preservation-rate}[ij] = 
\frac{1}{|\text{neighbors}[i]|} 
\sum_{k \in \text{neighbors}[i]} 
\frac{\sum_{l \in \text{neighbors}[j]} \mathbf{P}_{kl}}
     {\sum_{l=1}^{n_2} \mathbf{P}_{kl}}.
\]
The neighorhoods are defined as spots within Euclidean distance $\leq r$ including the spot itself, and neighbors of spots in slice A only contains spots in $V$ with non-zero mapping.

\noindent Finally, the global geometric preservation rate (GPR) of the alignment matrix $\mathbf{P}$ with neighborhoods defined with radius $r$ is obtained by averaging the preservation 
rates across all valid spots:  
\begin{equation}
\text{GPR}(\mathbf{P},r) = 
\frac{1}{|V|} \sum_{i \in V} \text{preservation-rate}[i].
\end{equation}

\noindent This metric captures how well the alignment preserves local neighborhood structures, 
providing a complementary view to accuracy by directly 
quantifying the geometric consistency of the alignment.

\subsection*{Cell-cell communication analysis}
In the cell-cell communication (CCC) analysis of DLPFC data, ligand-receptor pairs with known interactions through secreted ligands are obtained from CellChatDB \cite{jin2021inference}. Ligands and receptor genes expressed in less than 1\% of spots are filtered out. Then CCC analysis is performed using COMMOT \cite{cang2023screening} with a spatial distance cutoff of 500 $\mu$m. When comparing the spatial patterns of CCC across aligned slices, we only consider ligand-receptor (LR) pairs with more than 10\% of spots with non-zero received signal and non-zero sent signal. Consider slice A with $n_1$ spots and slice B with $n_2$ spots. Given a COMMOT communication matrix $\mathbf{W}_{\mathrm{A}}\in\mathbb{R}^{n_1\times n_1}$ of a LR pair, the receiver field and sender field in A are obtained by $\mathbf{R}_\mathrm{A}(j) = \sum_i \mathbf{W}_{\mathrm{A}}(i,j)$ and $\mathbf{S}_\mathrm{A}(i) = \sum_j \mathbf{W}_{\mathrm{A}}(i,j)$, respectively. The field, the example $\mathbf{R}_\mathrm{A}$, is mapped to slice B through the RAFT-UP mapping $\mathbf{P}\in\mathbb{R}^{n_1\times n_2}$ by $\mathbf{R}_{\mathrm{A}\rightarrow\mathrm{B}}=\bar{\mathbf{P}}^{\mathsf{T}}\mathbf{R}_\mathrm{A}$, where $\bar{\mathbf{P}}$ is the row-normalized $\mathbf{P}$.

\section*{Data availability}

The human DLPFC data with annotations \cite{pardo2022spatiallibd} is downloaded from github \url{https://benchmarkst-reproducibility.readthedocs.io/en/latest/Data%20availability.html} and zenodo \url{https://zenodo.org/records/10698880}, and the original data is available at \url{https://research.libd.org/spatialLIBD/}. The MERFISH data with annotations \cite{moffitt2018molecular} is downloaded from zenodo \url{https://zenodo.org/records/10698909} and the original data is available at \url{https://datadryad.org/dataset/doi:10.5061/dryad.8t8s248}. The Stereo-seq mouse midbrain at E12.5,E14.5 and E16.5 time points \cite{chen2022spatiotemporal} is downloaded from \url{https://spatrack.readthedocs.io/en/latest/notebooks/04.ST_data_of_mouse%20midbrain_with_multiple_times.html} and the original data is available at \url{https://db.cngb.org/stomics/datasets/STDS0000058/data}. The curated ligand-receptor pairs are taken from CellChatDB \cite{jin2021inference} available at \url{https://github.com/sqjin/CellChat}.

\section*{Code availability}

RAFT-UP is available as an open-source Python package at \url{https://github.com/L1feiyu/raftup_repo}.

\section*{Acknowledgement}

This work was supported by NSF grant DMS2142500 (Y.Z.), NSF grant DMS2151934 (Z.C.) and NIH grant R01GM152494 (Z.C.).

% \section*{Author Contributions}

% Y.W., Y.Z., and Z.C. conceived the method. Y.W., J.W., and Z.C. implemented the method and generated the numerical results. All authors interpreted the results, generated the visualizations and wrote the paper.

\section*{Competing Interests}
The authors declare no competing interests.

\nolinenumbers

\bibliography{sn-bibliography}

\end{document}